# Terrestrial Impact from the Passage of the Solar System through a Cold Cloud a Few Million Years Ago

**Authors:** Merav Opher[*1,2], Abraham Loeb[3]

**Affiliations:**

[*1] Radcliffe Institute for Advanced Studies at Harvard University, Cambridge, MA,

[2] Astronomy Department, Boston University, Boston, MA, USA

[3] Astronomy Department, Harvard University, Cambridge, MA, USA

*Correspondence to: Merav Opher, mopher@bu.edu

**Abstarct:** It is expected that as the Sun travels through the interstellar medium (ISM), there will be different filtration of Galactic Cosmic Rays (GCR) that affect Earth. The effect of GCR on Earth's atmosphere and climate is still uncertain. Although the interaction with molecular clouds was previously considered, the terrestrial impact of compact cold clouds was neglected. There is overwhelming geological evidence from $^{60}$Fe and $^{244}$Pu isotopes that Earth was in direct contact with the ISM 2-3 million years ago, and the local ISM is home to several nearby cold clouds. Here we show, with a state-of the art simulation that incorporate all the current knowledge about the heliosphere that if the solar system passed through a cloud such as Local Leo Cold Cloud, then the heliosphere which protects the solar system from interstellar particles, must have shrunk to a scale smaller than the Earth's orbit around the Sun (0.22AU). Using a magnetohydrodynamic simulation that includes charge exchange between neutral atoms and ions, we show that during the heliosphere shrinkage, Earth was exposed to a neutral hydrogen density of up to 3000cm$^{-3}$. This could have had drastic effects on Earth's climate and potentially on human evolution at that time, as suggested by existing data.

**Introduction:**

Most stars generate winds and move through the interstellar medium that surrounds them. This motion creates a cocoon that protects the close-in planets from the interstellar medium. We label these 'cocoons' astrospheres. The Sun's cocoon is the heliosphere. The solar system has been located inside a local super bubble (LB) for at least the last 3Myr and possibly 10Myr[27]. The conditions near the Sun are not homogeneous and several partially ionized clouds exist[16]. It is clear that the solar system has traversed different regions of the local interstellar medium (ISM) during the past several million years that have affected its heliosphere. Presently the solar system is traversing a local interstellar bubble with a relative speed of 25.7km/s. The solar system will be leaving it in the next few thousands of years because of its proximity to the bubble's edge[28].

Through geological radio isotopes on Earth we can learn about the past of the heliosphere. $^{60}$Fe has a half-life of 2.6 million years and $^{244}$Pu has a half-life of 80.7 million years. $^{60}$Fe is not naturally produced on Earth, and so its presence is an indicator of supernova explosions within the last few million years - being trapped in interstellar dust grains. $^{60}$Fe is predominantly produced in the winds of massive stars and in supernova explosions[29]. $^{244}$Pu is produced through the r-process thought to occur in neutron stars mergers[30]. Evidence of deposition of extraterrestrial $^{60}$Fe on earth was found in deep sea sediments and ferromanganese crusts between 1.7-3.2 million years ago (Ma)[9-12], in Antarctic snow[13] and in lunar samples[14]. The abundances were derived from new high precision accelerator mass spectrometry measurements. The $^{244}$Pu/$^{60}$Fe influx ratios are similar at ~2 Ma and a there is evidence of a second peak at ~7Ma[9-10]. In addition, cosmic ray data assembled by the Advanced Composition Explorer (ACE) spacecraft measured the $^{60}$Fe abundance as well[15]. This study estimated the time required for transport to Earth not to exceed the half time of $^{60}$Fe 2.6 million years and concluded that the cosmic rays diffused from a source closer than a distance of kpc. Studies have attributed the two peaks in $^{60}$Fe to multiple supernova explosions within 100pc over the last 10Myr that formed the local bubble[10] an processes that brought $^{244}$Pu to Earth through supernova ejecta or encounter of the Solar system with clouds enriched with $^{60}$Fe dust.

Other studies suggested that nearby supernova explosions within ~10-20pc could have produced the above isotopes[31]. In particular, the heliosphere would shrink down to just beyond 1AU for a nearby supernova as close as 10 pc. This scenario requires fine tuning since this distance is very close to the so called "kill radius" of 8pc[31], where extinction of all terrestrial life would have been triggered. For a supernova at a larger distance, there is a need to deliver $^{60}$Fe to Earth. It is also unclear whether traversing the shock width from a single supernova explosion would be consistent with the duration of 1.5Myr inferred from $^{60}$Fe data.

There have been some previous works on speculating that encounters of the heliosphere with molecular clouds could affect Earth environment[7,8]. Here we show that in the ISM that the sun has traversed within the last couple of Myr there are candidates for encounter with cold compact clouds that can drastically affect the evolution on Earth. It indicates that solar – like stars in common ISM will have major changes throughout their history.

Here we explore a different scenario whereby the solar system went through a cold gas cloud a few Myr ago. The ISM experienced today by the heliosphere is of a partially warm ionized medium with a hydrogen number density of $n_H \sim 0.2$cm$^{-3}$ and a temperature of $T \sim 8000$K[28]. The ISM in the vicinity of the solar system also harbors dense cold clouds. One of them is Local Leo Cold Cloud (LLCC)[32], thought to be part of a large ribbon of cold clouds[33]. The LLCC properties are estimated to be $n_H \sim 3000$cm$^3$ and $T=20$K[34]. Its distance was estimated first as 11-23 pc[35] and most recently as 45pc[36].

It is very possible that the heliosphere encountered 2Myrs ago a cold cloud like LLCC. The Sun moves through the Local Standard of Rest (LSR) in Galactic coordinates (U,V,W) with a velocity of (11.9,11.9,6.4) km s$^{-1}$ [37] or a net speed of 18.5pc/Myr. In 2Myrs, the Sun would have traversed 37pc with respect to the LSR. ISM clouds move with speeds of order tens of pc/Myr [16]. Using a Cartesian coordinate system centered on the Sun where the x-axis connects the Sun to the Galactic center, the y-axis is perpendicular to the x axis in the Galactic plane and the z-axis is perpendicular to the Galactic plane, the Sun is today at (x, y, z) = (0,0,0) pc and 2Myrs ago was at (-24.35,-24.35,-13.09) pc. Consider an example of a possible cold cloud at the LLCC location today of galactic longitude l=223°; and latitude b=44° [34]. Assuming a distance from the Sun of 45pc[34], LLCC is located today at (-23.67, -22.08, 31.26) pc.

The past kinematic properties of the LLCC are not well known, as it is connected to a much larger ribbon of cold clouds[33] that extends across the sky and could be the collision product of two clouds that previously approached each other[35]. In order to for the LLCC to collide with the Earth 2Myr ago, it had to have a velocity (U, V, W) = (-0.2, 0.7, -14.4) km/s with a magnitude of 14.5 km/s. This speed is typical of ISM clouds near the Sun[16], making the collision scenario plausible.

We simulate the interaction with of the heliosphere with a cold cloud like the LLCC 2Myr ago. The heliosphere's distance is currently ~ 130AU as measured by Voyager 1 and 2[38]. As our simulation demonstrates, the momentum deposition by the large hydrogen density of the cloud shrinks the heliosphere to a scale that is much smaller than the Earth's orbit around the Sun and brings the Earth and the Moon in direct contact with the cold ISM. Such an event must have had a dramatic impact on the Earth's climate.

**Results:**

Our computational code considers a single ionized component and four neutral components[17], although for this run we only used the ISM component that is orders of magnitude more abundant than the HS and supersonic solar wind components. The inner boundary is placed at 0.1 AU (or 21.5 solar radii). The adopted parameters of the solar wind at the inner boundary are based on the well benchmarked Alfven driven solar wind solution[39]. The grid is highly resolved at $1.07 \times 10^{-3}$ AU near the inner boundary and $4.6 \times 10^{-3}$ AU in the region of interest including the tail– see Methods – Figure 1. The run was performed for 1.3 years because our characteristic gas speeds of 50-100km/s traverse the needed scales of 9AU – 18AU during this time (see Methods for a description of the coordinate system, grid and model details). For the ISM outside the heliosphere we adopt the characteristics of LLCC[34], namely: $n_H$=3000cm$^3$; T=20K. We include a negligible ionized component ($n_i$=0.01; T=1K) and ignore the presence of interstellar magnetic field since its pressure is negligible compared to the ram pressure of the cold cloud. We adopt the relative speed such that the cold cloud crosses the path of the Sun 2Myr ago based on the current position of LLCC (see Methods). The relative velocity in Galactic coordinates between the LLCC and the Sun is $(\Delta U, \Delta V, \Delta W) = (-12.1, -12.6, 20.9)$km/s or a magnitude of 27.3 km/s. The neutral H from the cold cloud impinge on the heliosphere with speed of $U_x$ = 6.73 km/s; $U_y$= 26.33 km/s; $U_z$= -2.63 km/s. We rotate the system so the flow is in zx plane with the ISM approaches from the -x direction with $U_x$ = 27.18; $U_y$= 0; $U_z$= -2.63 km/s (see Methods for details).

The numerical model includes charge exchange between the neutrals and ions[17], as well the Sun's gravity which plays an important role of focusing the gas flow. We neglect radiation pressure from the Ly$\alpha$ line of hydrogen atoms since these cold dense clouds are optically-thick to Ly$\alpha$ photons[7]. We neglect photoionization, since its contribution is an order of magnitude smaller than that of charge exchange at these distances (see Methods for details).

Figure 1 shows the heliosphere as a result of the interaction with the cold cloud 2Myr ago. The heliosphere shrinks to $0.22 \pm 0.01$AU well within the Earth's orbit, exposing the Earth (and all other solar system planets for most of their trajectories to the ISM with neutral densities of 3000cm$^{-3}$ (Figure 1B). The supersonic solar wind goes through a termination shock (Figure 2A) before reaching equilibrium with the cold cloud. The heliosphere has a cometary shape with a long tail (Figure 2). Due to their small mean-free-path (~0.01-0.1AU), the neutrals get depleted quickly across the heliopause (Figure 3B) setting a strong gradient of ram pressure. The heliosphere reaches equilibrium with the cold cloud at the heliopause (HP) between the solar magnetic field compressed and the ram pressure of neutrals ahead of the HP (Methods - Figure 2). Gravity increases the density of neutrals from 3000cm$^{-3}$ and speed of 27.7km/s at large distances to 6110 cm$^{-3}$ and speeds of 88km/s near the HP (see Figure 2A; 3C and Methods - Figure 3).

The heliosphere 2Myr ago is very different that the heliosphere today[17]. There is no hydrogen wall since the number of ions ahead of the heliosphere is negligible. The heliosphere is so close to the Sun that the solar magnetic field is radial and the heliosheath plasma confinement does not take place[17]. The flow in the heliosheath (HS) is fast (~110-140km/s) (see Figure 2A and Figure 3D) and the ram pressure is larger than the magnetic pressure (Figure 2E). The Rayleigh-Taylor like instability that occurs currently in the HS[18] and drives the current heliosphere to have a short tail, is absent. Because of the short mean-free-path, there are almost no neutrals inside the heliosphere and the density gradient in the HS is absent as well. The Termination Shock (TS) shifts to distances as close as 0.14AU from the Sun. The present-day TS is weakened by pickup ions compared to the much stronger compression ratio of 3.7 of the TS during the passage of the cold cloud. This has consequences for acceleration of particles to high energies. We expect that the stronger shock accelerated GeV particles more efficiently than the current TS which is mediated by the presence of pickup ions[40]. Future work is needed to explore the resulting non-thermal emission and its consequences for the planets around the Sun or other stars.

Figure 2A shows the elongated tail filled with high speed velocity in the HS of the ancient heliosphere. Such elongated tails would be common for solar mass stars just born in dense interstellar environments like molecular clouds, and might have been misinterpreted in the past as jets.

**Discussion:**

The consequences of the Earth being exposed to pristine ISM are significant. For one, the encounter with the cold cloud could have shaped human evolution. This topic has received recent attention considering faunal and paleoclimate evidence exploring the hypothesis that past climate changes have influenced our evolution[21,22]. The period between 2-3Myr coincides with the extinction of *Australopithecus afarensis* (*"Lucy"*); the emergence of robust austrapoliths (*Paramthropus spp.*) *and the* emergence of the *Homo* lineage near time when the first evidence of stone tool manufacturing and transport appear[21].

This period contains the earliest appearance of *Homo erectus* the first hominin species to resemble modern humans and the first exodus out of Africa and into Europe and South Asia. The current view is that "African fauna, including our forebears may have been shaped by changes in climate variability. These views posit that increasing climate variability led to climate and ecological shifts that were progressive larger in amplitude"[21]. Environmental drivers have been attributed to hydroclimate extremes (arid or moist conditions) and habitat variability[23]. In particular, "deep-sea drilling of marine sedimentary sequences near the continent has recovered several continuous and well-dated records of wind-borne (eolian) dust variability," indicate a shift to a more arid climate between 2-3Myr[24].

This period of human evolution was the period where the environment changed, including cooling and wider climate fluctuations derived from the record of oxygen isotopes ($^{18}$O and $^{16}$O) over time. The oxygen record was measured in the microscopic skeletons of foraminifera in the sea floor[25]. Terrestrial and marine paleoclimate records have been interpreted to show that subtropical African climate has, over the past 2-3 Myr ago, progressively become more arid with an expansion of grasslands (ref *22, Page 25*). During that time the isotope data show an overall larger decrease in temperature and large degree of climate fluctuations, in particular during the later portion of human evolution (ref *22-Fig. 2.4*). The hypothesis is that the emergency of our species Homo sapiens was shaped by the need to adapt to climate changes at the time[26]. Large amounts of neutral hydrogen as a result of an encounter with cold clouds with densities above 1000cm$^{-3}$ was shown to alter the chemistry of Earth' atmosphere depleting ozone and creating global ice sheets that will have the final consequence of cooling Earth[19,20]. This cooling is aligned with what is seen in oxygen measurements in the sea floor[25].

We postulate that the encounter of the heliosphere with the cold cloud triggering the climate change was a critical component of the human evolution as a result of geographical migration. This hypothesis should be investigated with detailed climate models.

The extent of the signatures seen in $^{60}$Fe (higher time resolution that of the Pu signature) ~ 1.5Myr[9,10] is also consistent with an encounter with a cold cloud. Considering the relative speed and the typical cloud size of 10-20pc, it takes about 1.5Myr for the Sun to cross such a cloud. Our proposed scenario implies that all planets in the Solar system were exposed to the ISM simultaneously. It does not require absorption of $^{60}$Fe and $^{244}$Pu into dust particles which deliver them specifically to Earth, like the scenario of nearby Supernova explosions[10,31].

Another important effect on climate and evolution stems from Galactic Cosmic Rays (GCRs). Voyager 1 and 2 showed that the heliosphere shields the GCR for intensities < 200MeV by 75%[38]. During the passage through a cold cloud, Earth is exposed to the bare GCR intensity, enhanced further by the compression of the cold cloud if the GCR are trapped within the cloud. Detailed modeling of GCR diffusion is needed to explore the GCR impact on climate and habitability. For pedagogic purposes, we considered the LLCC as a generic example of a nearby cold cloud, and future observational studies could search for other nearby cold clouds that could have crossed the Solar system 2Myrs ago, as indicated by the $^{60}$Fe and $^{244}$Pu data.

**Methods**
**Description of the Numerical Model:**

The cold thermal solar wind and hot pickup ions (PUIs) are treated as a single species. The neutral hydrogen component is captured with a four-fluid approximation[41,42] although for this problem only the supersonic component and the cold pristine ISM take part in the interaction since there is no shock in the solar wind as the termination shock or a bow shock in the ISM as it approaches the heliosphere.

We neglect radiation pressure from the Ly$\alpha$ line of hydrogen atoms since these cold dense clouds are optically thick to Ly$\alpha$ photons with an optical depth $\tau$ much larger than unity: at the column density of hydrogen atoms in LLCC: N ~ [10$^3$ cm$^{-3}$]*[pc] ~ 10$^{21}$ cm$^{-2}$. The Ly$\alpha$ cross-section at resonance is $\sigma$~7x10$^{-11}$ cm$^2$ (*ref.44*) and so $\tau$~N$\sigma$~10$^{11}$. Hence, radiation was inferred to play a smaller role than gravity (same as in *ref. 7*). This is different than in current ISM conditions where radiation pressure is comparable the gravity[45]. We neglect photoionization since its contribution is an order of magnitude smaller than that of charge exchange at these distances.

The computational model coordinate system is such that the z axis is parallel to the solar rotation axis, the x axis is oriented in the direction of the interstellar flow (which points 5° upward in the x–z plane) and the y axis completes the right-handed coordinate system where the Sun is at rest at the center.

Future work can explore the current scenario with more advanced codes where the solar wind ions are treated as separate components[46] or the neutral hydrogen atoms are treated kinetically[47]. We do not expect though any major changes from the current results. The density of neutrals is so high that a fluid treatment is appropriate[7]. The separation of thermal and suprathermals will further enhance our results bringing further-in the heliosphere. This is because the the pick-up ions (PUIs) charge exchange (the mean free path for keV PUIs ~ $0.01 AU^{-1}$ (for densities as high as $10^3 cm^{-3}$) and leave the system deflating the heliosphere.

**Inner boundary**:

The inner boundary is placed at 0.1 AU (or 21.5 solar radii). The adopted parameters of the solar wind at the inner boundary are: $v_{SW}$ = 417 km/s, $n_{SW}$ = 5.71 x $10^2$ cm$^{-3}$, $T_{SW}$ = 2.59 x $10^5$ K based on the Alfven driven solar wind solution[39]. The magnetic field is given by the Parker spiral magnetic field[43] with $B_{SW}$ = 1.72x$10^2$ nT at the equator. We use a monopole configuration for the solar magnetic field (as in refs. 17, 48). This description, while capturing the topology of the field lines, does not capture the change of polarity with solar cycle or across the heliospheric current sheet. This choice, however, minimizes artificial reconnection effects, especially in the heliospheric current sheet. We assume that the magnetic axis is aligned with the solar rotation axis.

**ISM conditions**:

Neutral hydrogen atoms $n_H$ = 3000.0 cm$^{-3}$, $T_H$ = 20 K, $n_{ISM}$ = 0.01 cm$^{-3}$ and $T_{ISM}$ = 1 K. Both were streaming with speed in the coordinate system of the model with Ux=6.71km/s; Uy=26.27km/s; -2.63km/s (see Description of Coordinate Transformation between galactic to model coordinates below).

**Grid resolution**:

The grid extends $\pm$50AU in y and z and -20-50AU in x. We cover all regions of interest with high grid resolution (minimum grid cell is 1.07x$10^{-3}$ AU near the inner boundary and 4.6 x$10^{-3}$ AU in the region of interest including the tail– see Methods - Fig. 1. The simulation was run to 1.3 years. This time is appropriate because for speeds as 50-100km/s as in this simulation in a year reach 9AU – 18AU.

For Methods-Figure 3, the run with no gravity, the inner boundary and ISM conditions were the same as the run with gravity. The grid was smaller being $\pm$20AU in x, y, z with resolution of 1.07x$10^{-3}$ AU near the inner boundary, 4.07x$10^{-3}$ AU within the Termination Shock and 0.03AU in the Heliosheath. Since we were only interested in showing the difference at the nose we run it for a shorter time to 0.1year.

**Description of the Coordinate Transformation between Galactic Coordinates to the Model Coordinates.**

The relative velocities between the Sun and the Cold Cloud in galactic coordinates are $U_x$=-12.1, $U_y$=-12.6, $U_z$=20.9 km/s which correspond to galactic coordinates: (lat, lon)=(570.11°,-133.84°). Converting galactic to ecliptic coordinates (https://lambda.gsfc.nasa.gov/toolbox/tb_coordconv.cfm) for J2000 epoch this corresponds to latitude and longitude in Ecliptic Coordinates: (lat, lon)=(1.46°, 150.48°).

In HCI coordinate system (our model) this correspond to (lat, lon)=(-5.54°,75.67°) or to coordinate vector: (x_HCI, y_HCI, z_HCI)=(0.2464,0.9644,-0.0965). This corresponds to the relative speeds in HCI of $U_{x\_HCI}$ = 6.71 km/s; $U_{y\_HCI}$ = 26.27 km/s; $U_{z\_HCI}$ = -2.63 km/s.

**References**

1. Frisch, P. C., & Muller, H.-R. Time-Variability in the Interstellar Boundary Conditions of the Heliosphere: Effect of the Solar Journey on the Galactic Cosmic Ray Flux at Earth. *Space Science Reviews* **176**, 21-34 (2013).

2. Muller, H.R, Frisch, P. C., Florisnki, V., and Zank, G. P. Heliospheric Response to Different Possible Interstellar Enviornment. *Astrophys. J.* **647**, 1491-1505 (2006).

3. Florinski, V., Zank, G. P., and Axford, W. I. The Solar System in a dense interstellar cloud: Implications for cosmic-ray fluxes at Earth and $^{10}$Be records. *Geophys. Res. Letters* **30**, No. 23, 1-5 (2003).

4. Scherer, K., Fichtner, H., Heber, B. et al. Cosmic ray flux at the Earth in a variable heliosphere, *Advances in Space Research* **41**, 1171-1176 (2008).

5. Shaviv, N. J. Cosmic Ray Diffusion from the Galactic Spiral Arms, Iron Meteorites, and a possible climate connection? *Phys. Rev. Lett.* **89**, 051102-051104 (2002).

6. Kikby, J. & Carslaw, K. S., Variation of Galactic Cosmic Rays and the Earth Climate, Chapter 12 of Solar Journey, the Significance of Our Galactic Environment for the Heliosphere and Earth, Pricilla C. Frisch Editor, Springer (2006).

7. Yeghikyan, A. and Fahr, H., Chapter 11 of Solar Journey, the Significance of Our Galactic Environment for the Heliosphere and Earth, Pricilla C. Frisch Editor, Springer (2006).

8. Begelman, M. C. & Rees, M. J. Can cosmic clouds cause climatic catastrophes? *Nature* **261**, 298-299 (1976).

9. Wallner, A. et al. Recent near-Earth supernovae probed by global deposition of interstellar radioactive $^{60}$Fe. *Nature* **532**, 69-72 (2016).

10. Wallner, A. et al. $^{60}$Fe and $^{244}$Pu deposited on Earth constrain the r-process yields of recent nearby supernovae. *Science* **372**, 742-745 (2021).

11. Knie, K. et al. Indication for Supernova Produced $^{60}$Fe Activity on Earth. *Phys. Rev. Lett.* **83**, 18-21 (1999).

12. Fitoussi, C. et al. Search for Supernova-Produced $^{60}$Fe in a Marine Sediment. *Phys. Rev. Lett.* **101**, 121101 (2008).

13. Koll, D. et al. Interstellar $^{60}$Fe in Antarctica. *Phys. Rev. Lett.* **123**, 072701 (2019).

14. Fimiani, L. et al. Interstellar $^{60}$Fe on the Surface of the Moon. *Phys. Rev. Lett.* **116**, 151104 (2016).

15. Binns, W. R. et al. Observation of the $^{60}$Fe nucleosynthesis-clock isotope in galactic cosmic rays. *Science,* **352**, 677-680 (2016).

16. Redfield, S., & Linsky, J. L. The Structure of the Local Interstellar Medium. IV. Dynamics, Morphology, Physical Properties, and Implications of Cloud-Cloud Interactions. *Astrophys. J.* **673**, 283-314 (2008).

17. Opher, M., Drake, J. F., Zieger, B., Gombosi, T. I. Magnetized Jets Driven by the Sun: The Structure of the Heliosphere Revisited, *The Astrophysical Journal Letters,* **800**, L28 (19pp) (2015).

18. Opher, M., Drake, J. F., Zank., et al. A Turbulent Heliosheath Driven by Rayleigh Taylor Instability, *The Astrophysical Journal*, **922**, 181 (12pp) (2021).

19. McKay, C. and Thomas, G. E., Consequence of a Past Encounter of the Earth with an Interstellar Cloud. *Geophys. Res. Lett.* **5**, No. 3, 215-218 (1978).

20. Yabushita, S. and Allen, A. Did an Impact alone kill the dinosaurs? *Astronomy & Geophysics* **38**, Issue 2, 15-19 (1997).

21. DeMenocal, P. B. Climate and Human Evolution, *Science* **331**, 540-542 (2011).

22. National Research Council, Understanding Climate's Influence on Human Evolution (National Academies Press, Washington, DC, 2010).

23. Potts, R. and Faith. J. T. Alternating high and low climate variability: The context of natural selection and speciation in Plio-Pleistocene hominin evolution *J. Hum. Evol.* **87**, 5-20 (2015).

24. deMenocal, P. B. Plio-Pleistocene African Climate. *Science* **270**, 53-59 (1995).

25. Zachos, J. et al. Trends, Rhytms, and Aberrations in Global Climate 65 Ma to Present, *Science* **292**, 686-693 (2001).

26. Potts, R. et al. Environmental dynamics during the onset of the Middle Stone Age in eastern Africa., *Science* **360**, 86-90 (2018).

27. Fuchs, B. et al. The search for the origin of the Local Bubble redivivus. *Mon. Not. R. Astron. Soc.* **373**, 993-1003 (2006).

28. Frisch, P. C. et al. The Galactic Environment of the Sun: Interstellar Material Inside and Outside of the Heliosphere. *Space Sci. Rev.* **146**, 235-273 (2009).

29. Wang, W. et al. SPI observations of the diffuse $^{60}$Fe emission in the Galaxy. *Astron. Astrophys.* **469**, 1005-1012 (2007).
8

30. Ji., A. P. et al. R-process enrichment from a single event in an ancient dwarf galaxy. *Nature* **531**, 610-613 (2016).

31. Field, B., Athanassiadou, T., Johnson, S. R., Supernova collisions with the heliosphere. *Astrophys. J.* **678**, 549-562 (2008).

32. Meyer, D. M., et al. A Cold Nearby Cloud Inside the Local Bubble. *Astrophys. J.* 650, L67-L70 (2006).

33. Haud, U. Gaussian decomposition of HI surveys, *A&A* **514**, A27-A34 (2010).

34. Meyer, D. M., Lauroesch, J. T., Peek, J. E. G., Heiles. C., *Astrophys. J.* **752**, 119 (15pp) (2012).

35. Peek, J. E. G. et al. The Local leo Cold Cloud and New Limits on a Local Hot Bubble/ *Astrophys. J.* **735**, 129 (12pp) (2011).

36. Gry, C., Jenkins, E. B., The nearby interstellar medium toward Leo UV observations and modeling of a warm cloud within hot gas. *A&A* **598**, A31-A49 (2017).

37. Zbinden, O., Saha, P. *Res. Notes AAS* **3**, 73 (2019).

38. Stone, E. C. et al. Cosmic ray measurements from Voyager 2 as it crossed into interstellar space. *Nature Astronomy* **3**, 1013-1018 (2019).

39. Evans, R., Opher, M., Oran, R., et al., Coronal Heating by Surface Alfven Wave Damping: Implementation in a Global Magnetohydrodynamics Model of the Solar Wind, *The Astrophysical Journal*, **756**, Issue 2, article id. 155 (13pp) (2012).

40. Giacalone, J., et al. Hybrid simulations of Interstellar Pickup Protons Accelerated at the Solar-Wind Termination Shock at Multiple Location, *The Astrophysical Journal*, **911**, Issue 1, id. 27, 8pp (2021).

41. Zank, G. P. Interaction of the Solar Wind with the Local Interstellar Medium: a Theoretical Perspective. *Space Science Reviews* **89**, Issue 3/4, 413-688 (1999).
42. Opher, M., et al., A Strong highly-tilted Interstellar Magnetic Field near the Solar System, *Nature*, **462**, 1036-1038 (2009).
43. Parker, E. Dynamics of the Interplanetary Gas and Magnetic Fields. *Astrophys. J.* **128**, 664-676 (1958).
44. Lecture notes of Lyman alpha radiation transfer; https://arxiv.org/pdf/1704.03416.pdf
45. Schwadron, N. et al. Solar Radiation Pressure and Local Interstellar Medium Flow Parameters From Interstellar Boundary Explorer Low Energy Hydrogen Measurements. *Astrophys. J.* **775**, 86 (13pp) (2013).
46. Opher, M., Loeb, A. Drake, J., Gabor, T. A Predicted Small and Round Heliosphere suggested by magnetohydrodynamic modelling of pick-up ions, *Nature Astronomy,* **4**, 675-683 (2020).

47. Michael, A. T. et al. The Solar-wind with Hydrogen Ion Exchange and Large-Scale Dynamics (SHIELD) code: A Self-Consistent Kinetic-MHD Model of the Outer Heliosphere *Astrophys. J Supplements,* **924***, Number 2*, 105 (2022).
48. Izmodenov, V.V. & Alexashov, D. B., Three-Dimensional Kinetic-MHD Model of the Global Heliosphere with the Heliopause-Surface Fitting. *Astrophys. J Supplements Series*, **220**, Issue 2, article id. 32, 14 pp. (2015).

**Acknowledgments:** The authors would like to thank discussions with M. Kornbleuth and help on the coordinate conversion. This work is supported by NASA grant 18-DRIVE18_2-0029 as part of the NASA/DRIVE program titled "Our Heliospheric Shield". For more information about this center please visit: http://sites.bu.edu/shield-drive/. M.O. was supported as well by the Fellowship Program, Radcliffe Institute for Advanced Study at Harvard University.

**Author contributions:** Conceptualization; Methodology, Investigation, Interpretation and Conclusions, Writing: M.O. and A.L.; Numerical calculation and Visualization: M.O.

**Data and materials availability:** Our model is the outer heliosphere (OH) component of the SWMF and is available at http://csem.engin.umich.edu/tools/swmf/. The data produced by the model that support the findings of this study are available from the corresponding author upon reasonable request.

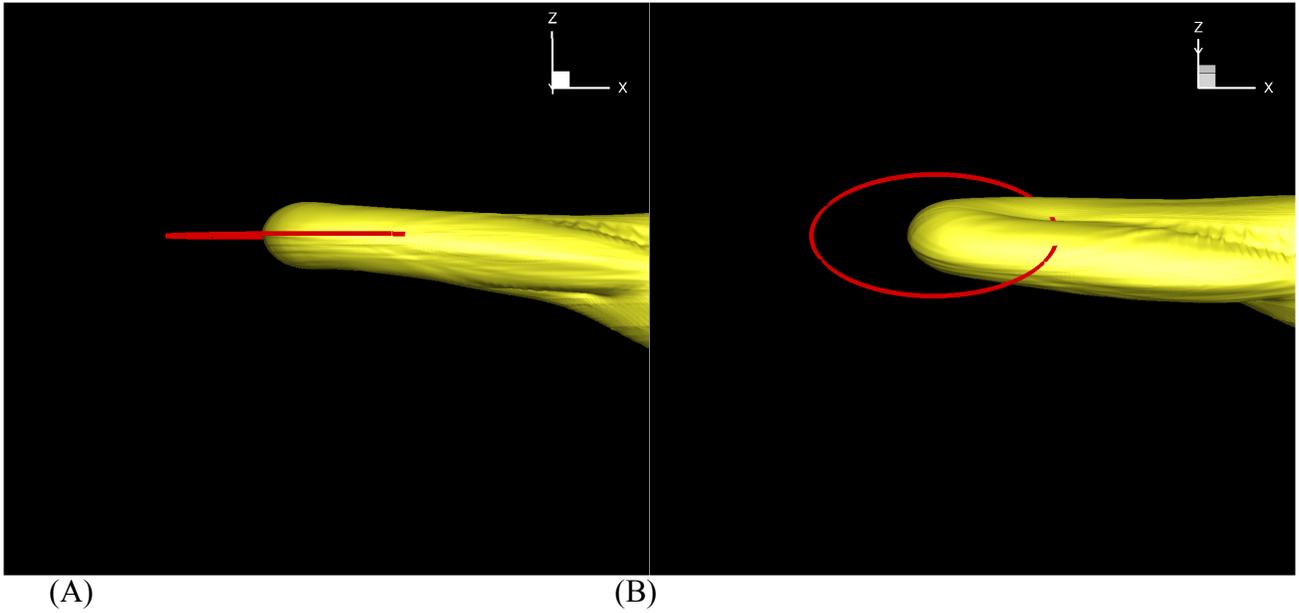

**Fig. 1. 3D image of heliosphere with two views**. The trajectory of Earth is plotted in red. Iso-surface of the heliosphere is plotted at neutral density $n_H=2000 cm^{-3}$. We plotted the tail out to 4AU.

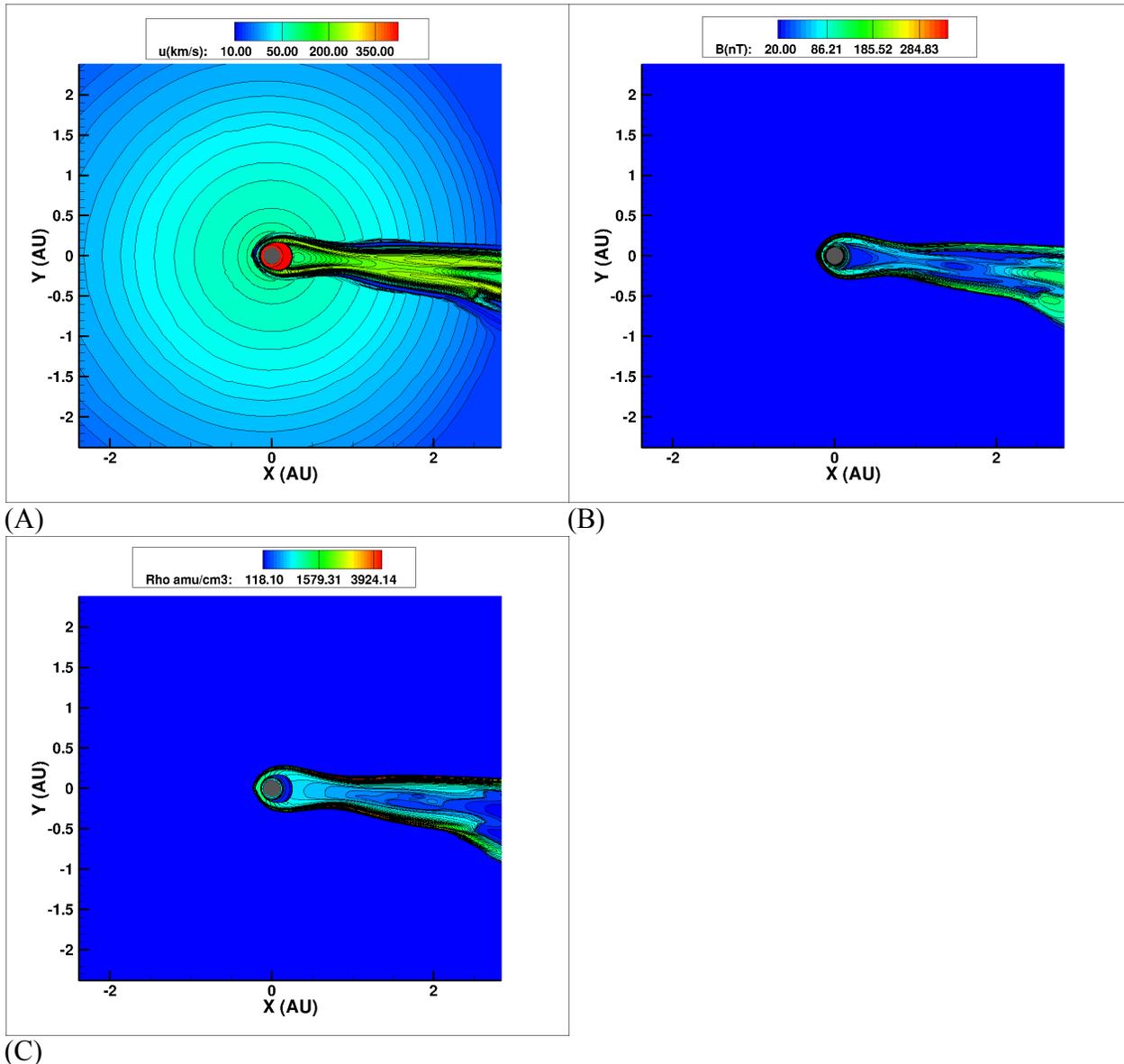

**Fig. 2. Heliosphere 2 Myr ago.** Panels are shown at the end of simulation at 1.3 yr. Panels (A)-(C) are in the meridional plane at y=0 AU (for the model coordinate system see SI). Contours are (A) speed; (B) magnetic field; and (C) ion density. The Heliosphere shrinks to 0.22AU at the nose, maintaining a long cometary shape and exposing all planets to the cold dense ISM material.

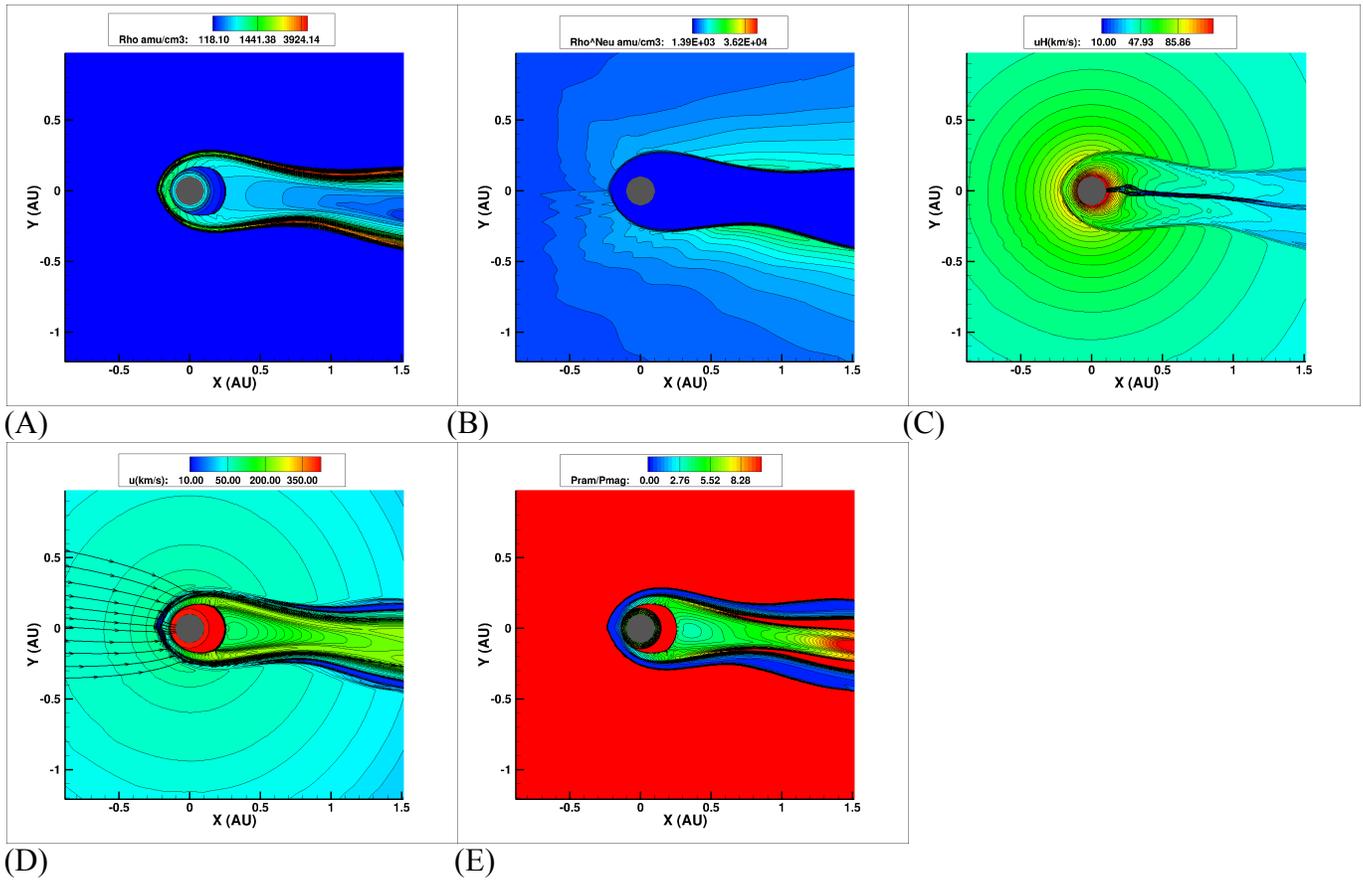

**Fig. 3. Heliosphere 2 Myr ago** – close view of Fig. 2. Panels are shown at the end of simulation at 1.3 yr. Panels (A)-(E) are in the meridional plane at y=0 AU (for the model coordinate system see SI). Contours are (A) ion density; (B) neutral density; (C) neutral speed; (D) ion speed; (E) Ratio of ram pressure to magnetic pressure.

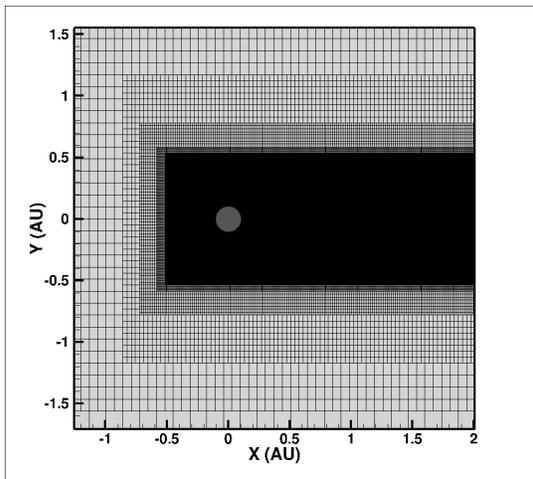

**Methods - Figure 1.** Grid is shown in a zoom region around the region of interest in the meridional plane at y=0.

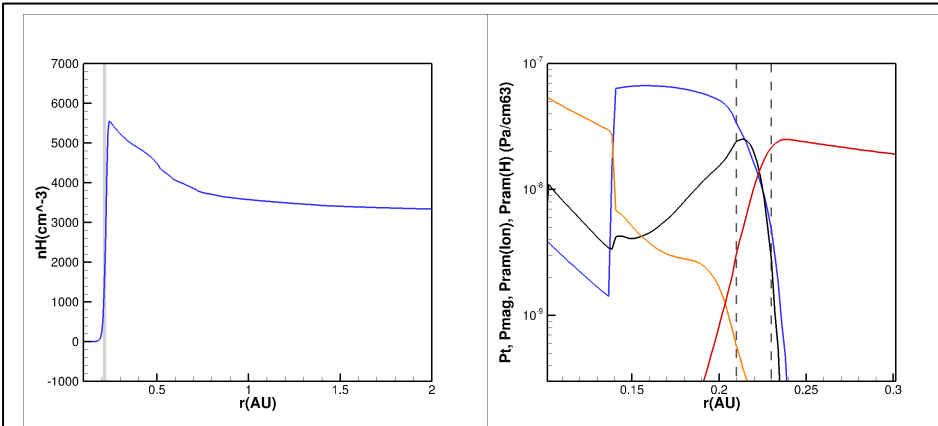

(a)                                      (b)

**Methods - Figure 2.** Upstream cut showing (a) the neutral density; and (b) pressures – blue line is the ion thermal pressure; black line the magnetic pressure; the red line the ram neutral pressure and the orange the ram ion pressure. The heliopause location is shown at $0.22 \pm 0.01$AU.

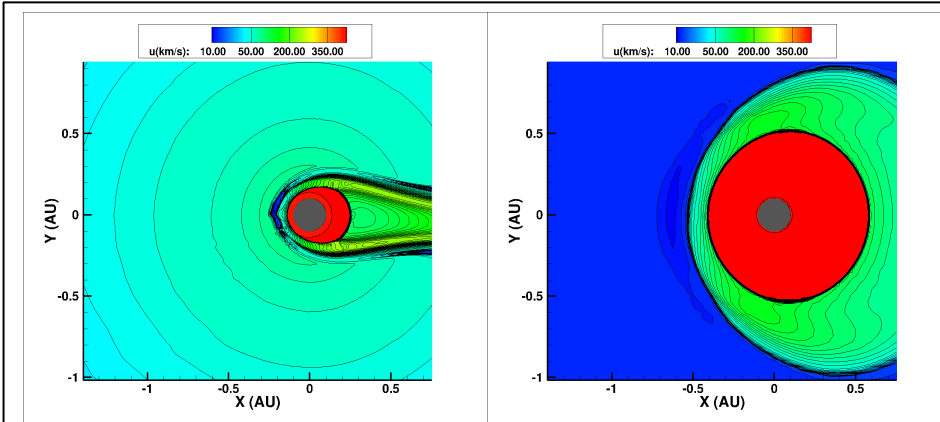

(a)                         (b)

**Methods - Figure 3.** Neutral speed at the nose in the meridional cut for the run (a) with Gravity; and (b) no gravity. All other conditions were the same.